# A 5.7 THz GaN/AlGaN quantum cascade detector based on polar step quantum wells

AFFILIATIONS

P. Quach,[1,a)] A. Jollivet,[1] A. Babichev,[2,3] N. Isac,[1] M. Morassi,[1] A. Lemaitre,[1] P. A. Yunin,[4] E. Frayssinet,[5] P. de Mierry,[5] M. Jeannin,[1] A. Bousseksou,[1] R. Colombelli,[1] M. Tchernycheva,[1] Y. Cordier,[5] and F. H. Julien[1]

[1]*Centre de Nanosciences et de Nanotechnologies (C2N), CNRS UMR 9001, University Paris-Saclay, 8 Avenue de la Vauve, 91120 Palaiseau, France*

[2]*ITMO University, 49 Kronverksky Pr., 197101 St. Petersburg, Russia*

[3]*Saint Petersburg Electrotechnical University "LETI", 5 Professora Popova, 197376 St. Petersburg, Russia*

[4]*Institute for Physics of Microstructures, RAS, GSP-105, 603950 Nizhny Novgorod, Russia*

[5]*Université Côte d'Azur, CNRS, CRHEA, Rue Bernard Grégory, 06560 Valbonne, France*

[a)]Author to whom correspondence should be addressed: patrick.quach.guoyongbin@gmail.com

ABSTRACT

We report on a GaN/AlGaN quantum cascade detector operating in the terahertz spectral range. The device was grown by metal organic chemical vapor deposition on a *c*-sapphire substrate and relies on polar GaN/AlGaN step quantum wells. The active region thickness is in micrometer range. The structural, electrical and optical investigations attest of high structural quality of the synthetized nitride material. The detector exhibits a peak photocurrent at 5.7 THz (23.6 meV) with a responsivity of 0.1 mA/W at 10 K under surface normal irradiation through a 10 μm period grating. The photocurrent persists up to 20 K.

GaN/Al(Ga)N quantum wells (QWs) are an interesting semiconductor material system for intersubband (ISB) devices. Their wide bandgap and high conduction band offset (1.75 eV for GaN/AlN)[1] allow operation across a wide spectral range, from the near infrared (NIR) to the terahertz (THz). NIR ISB absorptions in GaN/Al(Ga)N QWs require good structural quality synthesis with precise thickness and composition control. It was achieved in 2000s as a result of the development of the epitaxial growth of III-nitrides materials with metal organic chemical vapor deposition (MOCVD)[2,3] and molecular beam epitaxy (MBE).[4,5] These results opened the path to the realization of a vast number of different Al(Ga)N/GaN devices operating in NIR such as detectors,[6–8] all-optical switches,[9,10] electro-optical modulators,[11,12] pumped light emitters,[13,14] non-linear devices (second harmonic generation)[15,16] and ISB polaritons.[17,18] Recently, by specifically designing a quantum cascade detector (QCD) with active wells containing six bound states, we achieved an $e_1e_6$ ISB transition in the visible range.[19]

On the other side of the infrared spectrum, tuning ISB transitions at THz frequencies in AlGaN/GaN QWs is not straightforward. For polar III-nitride materials with a standard wurtzite structure grown along the c-axis, the presence of a strong internal electric field resulting from the polarization discontinuity between GaN and AlGaN along the c growth axis prevents tuning the ISB transitions into the THz range in simple QWs.[20] However, this issue can be overcome with a strategy relying on polar step-QWs grown along the standard *c*-axis orientation. These step-QW structures consist of a GaN well, an AlGaN barrier and an AlGaN step-layer with a lower aluminium (Al) concentration. Through a proper choice of the Al concentration, it is possible to minimize the internal field in the step-layer and to mimic a square potential. This strategy has proved effective using step-QWs composed of a GaN well, an $Al_{0.05}Ga_{0.95}N$ step layer, and an $Al_{0.1}Ga_{0.9}N$ barrier. THz range ISB absorptions have been in fact observed in AlGaN/GaN step-QWs synthetized by MBE[21] and MOCVD.[22] These step-QWs were also used to demonstrate a III-nitride-based quantum well infrared photodetector (QWIP) operating at 13 THz.[23] One of the many challenges in terms of material growth is to precisely control the Al concentration in the barriers to flatten the potential profile.

Alternatively, this internal electric field issue can also be overcome by growing alternate form non-polar III-nitride layers. Acceptable structural quality synthesis and observation of THz range ISB absorption was achieved with AlGaN/GaN QWs in the cubic form[24] as well as wurtzite structures grown along non-polar *m*-planes.[25,26] A III-nitride-based QWIP grown with a semi-polar orientation also showed detection at ~10 THz.[27]

An additional noticeable peculiarity of III-nitride materials is their large longitudinal optical (LO) phonon energy (92 meV in GaN and 110 meV in AlN). This property offers prospects towards THz quantum cascade lasers (QCLs)[28,29] operating above room temperature and giving access to a larger spectral range (1 to 15 THz), contrary to other III-V semiconductors. Two conditions are necessary to achieve demonstration of a GaN THz QCL. The first one is to achieve reliable resonant tunneling. QCLs[30] rely on resonant tunneling for efficient carrier injection from a period of the active region to the adjacent one. Continuous improvement of III-nitride epitaxy and cleanroom processing led in 2016 to the achievement of the first repeatable room temperature (RT) negative differential resistance (NDR) in an AlN/GaN resonant tunneling diode (RTD) with a peak-to-valley-current ratio (PVCR) of 1.15 at RT.[31] Later on, this result in AlN/GaN RTDs was observed and confirmed by several other groups.[32,33] Constant progress led to a recent demonstration of a PVCR of 1.24 at RT, and of

2.96 at 6.6 K.[34]

The second condition is the synthesis of high structural quality, micrometer-thick active regions that provide enough optical gain to achieve lasing. In THz QCLs specifically, the active region thickness must be of at least a few micrometers.[35,36] Consequently, a milestone step toward the development of III-nitrides THz QCL would be the realization of a III-nitrides quantum cascade device approaching this active region thickness and able to demonstrate vertical transport and ISB transitions in the THz range. A GaN/AlGaN THz QCD is a suitable tool to achieve this demonstration.

In this letter, we report on a successful monocrystalline growth of a GaN/AlGaN QCD with active region thickness in the micrometer range, able to show efficient vertical transport and operation in the THz spectral range. The device relies on polar GaN/AlGaN step QWs for the active and extractor regions and was grown by MOCVD on a *c*-sapphire substrate. Characterizations by X-ray diffraction and atomic force microscopy (AFM) reveal the high structural quality of the sample. The QCD exhibits a peak photocurrent at 5.7 THz (23.6 meV) with a calibrated responsivity of 0.1 mA/W at 10 K under normal incidence irradiation through a 10 µm period grating. In addition to the structural characterizations, observation of a Lorentzian lineshape with a reduced linewidth of 5.6 meV of the photocurrent spectra as well as the good agreement between experimental current-voltage (I-V) characteristics and theoretical prediction based on conduction band structure attest to the high structural quality of the sample.

Fig. 1 shows the conduction band profile of one period of the QCD calculated with the 8-band **k.p** Schrödinger-Poisson solver Nextnano.[37] The calculations were performed assuming periodic boundary conditions. A period is composed of five step-QWs consisting of an active well followed by a four-step-QWs-extractor. Each step-QW is made of a GaN well, followed by an $Al_{0.05}Ga_{0.95}N$ step layer, sandwiched between two $Al_{0.1}Ga_{0.9}N$ barriers. The layer sequence of one period of the structure, in nm, from left to right, starting from the active well is **3.9**/ 3.9/ $\overline{7.7}$/ **3.1**/ $\underline{3.1}$/ $\overline{1.2}$/ **3.1**/ $\underline{3.1}$/ $\overline{1.6}$/ **3.1**/ $\underline{3.1}$/ $\overline{2.0}$/ **3.1**/ $\underline{3.1}$/ $\overline{2.7}$ nm. $Al_{0.1}Ga_{0.9}N$ barriers are indicated in bold font and $Al_{0.05}Ga_{0.95}N$ step layers are indicated by the upperline. The underlined figures correspond to the GaN layers, n-doped with silicon at $n = 7 \times 10^{17}$ $cm^{-3}$. By using step-QWs, one can adjust the ISB transition energy by adjusting the $Al_{0.05}Ga_{0.95}N$ step layer thickness. As seen in Fig. 1, in the active step-QW, electrons from the ground state $e_1$ are photoexcited to the $e_2$ and $e_3$ states. The ISB transition energies are predicted at $E_{13}$ = 24.9 meV (50 μm or 6 THz) and $E_{12}$ = 31.1 meV (40 μm or 7.5 THz).

The electrons photoexcited into the $e_2$ or $e_3$ states relax through the extractor toward the next period. The whole photoexcitation and relaxation process occurs between bound states. Note: in the mid-infrared spectral range, the extraction of photo-electrons through the extractor multi-QW stages is mediated by ultrafast LO-phonon scattering between the extractor confined states.[38,39] In the THz spectral range,[40] the inter-level separation is too small with respect to the LO-phonon energy and other scattering mechanisms[41,42] are required for energy relaxation, such as interface roughness scattering (IRS), alloy disorder scattering (AD), ionized impurity scattering (II), electron-electron scattering (EE). Since IRS and AD are epitaxy related and difficult to control, we chose to use silicon doping in the extraction region to induce II and EE scattering in order to enhance the electron transfer efficiency to the next period.

For efficient operation of a QCD, the Fermi energy, which depends on the doping concentration and operating temperature, should be higher than the active QW ground state $e_1$ energy but lower than the last extractor state $e_6$ energy. The electron concentration also needs to be high enough to provide an ISB absorption able to generate a detectable photocurrent, but it needs to be low enough so to keep a high resistance and a low Johnson noise. Since in a THz QCD, the energy difference between the last extractor state $e_6$ and the ground state $e_1$ energy level is very small, the doping concentration window is narrow. In each extractor step-QWs of our device, the GaN well at the right of the $Al_{0.1}Ga_{0.9}N$ barrier is doped with Si at $7 \times 10^{17}$ $cm^{-3}$ corresponding to a Fermi level of $E_F - E_1 = 9.4$ meV above the active well ground state energy and $E_6 - E_F = 3.5$ meV below the lowest extractor states energy at 10 K.

The sample was grown by MOCVD on a 4-µm-thick GaN buffer on *c*-sapphire substrate. The 1-µm-thick active region contains a 20-period active structure sandwiched between an $Al_{0.05}Ga_{0.95}N$ cladding layer doped with Si at $1 \times 10^{17}$ $cm^{-3}$ as a bottom contact layer and a GaN cladding layer doped with Si at $2.5 \times 10^{18}$ $cm^{-3}$ as a top contact with a respective thickness of 1000 nm and 50 nm. The Al composition in the bottom cladding layer is close to the average Al content in the active region in order to reduce strain across the active region.

The grown sample was analyzed with X-ray diffraction measurements using a Cu source ($\lambda_{K\alpha}$=0.154056 nm). The high number of satellite peaks attests the good periodicity and structural quality of the structure (see Fig. 2). Based on the satellite peaks positions, the period thickness was deduced to be 46.94 nm, equal to the nominal value within a 2 % deviation. The best fit to the experimental measurement is achieved for Al concentrations of 4.5 % and 9 % which are close to the nominal values. Fig. 3 shows an AFM image of the sample surface which confirms the step-flow feature and smooth surface morphology of the as-grown samples. The root mean square roughness is smaller than 0.6 nm over an area of 5 × 5 μm².

After the epitaxy, the sample has been processed in the form of rectangular mesas with a size of 500 × 600 μm² using standard optical lithography followed by inductively coupled plasma reactive ion etching. Ti/Al/Ti/Au metallic layers have been deposited on the top and bottom contacts. The sapphire substrate being highly opaque in the THz range in contrast to silicon,[21,22] the observation of a THz ISB absorption under irradiation through a 45° angle polished facet is not possible. Therefore, light injection has to be done by the mesa surface and top contacts were patterned with metal grating to allow efficient coupling of incident light with ISB transition in the sample QWs. 10 μm, 16 μm and 20 μm period grating with 50 % duty cycle were fabricated. The contacts were annealed at 700°C for 60 s in order to prevent a Schottky barrier blocking of the photocurrent.

The dark current-voltage (I-V) characteristics have been measured from 10 K to 80 K, using a probe station and a Keithley 2636 SourceMeter. Fig. 4 represents the measured curves. From 50 K to 80 K, the dark current is dominated by the thermally-activated escape of carriers from the QWs to the continuum as attested by the ohmic I-V characteristics. With the temperature decreasing from 40 K down to 10 K, the I-V shows a more non-linear behavior. Figure 5 shows the measured IV characteristics at 10 K. We observe a transition around -0.75 V.

There are two regimes: from 0 V to -0.75 V, a non-linear characteristic is observed. It corresponds to a bias increase until alignment between two periods. While from -0.75 V to -1.5 V, the I-V characteristics become more linear. It shows an effective alignment between periods and injections of each period QWs bounded carriers to the adjacent period QWs bounded states. The inset of Fig. 5 shows the conduction band profile of the QCD under an applied bias of -0.73 V (corresponding to an electric field of -7.73 kV/cm and a voltage drop of -36 mV per period) at 10 K. At this specific bias, the ground state from the next period $e_1$, is aligned with the previous period state $e_4$ (the extractor second QW state). This agreement between theoretical conduction band QCD and sample measured I-V characteristics attests of the absence of strong parasitic conduction channels and of the high structural quality of the grown sample.

The photocurrent spectroscopy was performed using a Fourier transform IR (FTIR) spectrometer with a mechanically chopped glowbar light source and a lock-in amplifier. The QCD preliminary optical characterization at 0 V did not reveal any measurable photocurrent. We then applied a small bias in order to compensate for the band bending between the AlGaN active region and the GaN top cladding layers.[43] An optimum signal to noise ratio was found for an applied voltage of -50 mV since the noise level increased for higher and lower applied bias. Fig. 6 shows the photocurrent spectrum at 10 K under a bias of -50 mV for a QCD with a 10-μm period grating. Assuming a linear voltage drop, the voltage drop across one period is less than 2.5 mV, which is small compared with the transition energy of ~ 24.9 meV. The measured photocurrent spectrum is peaked at 5.7 THz (23.6 meV) with a shoulder around 7.5 THz (31 meV). This is in good agreement with the simulated transition energies $e_1e_3$ = 24.9 meV (6 THz) and $e_1e_2$ = 31.1 meV (7.5 THz) respectively. The $e_1e_2$ ISB transition peak intensity is much lower than that of the $e_1e_3$ peak since electrons photoexcited from $e_1$ to $e_2$ have a much higher probability to relax back to $e_1$ than to tunnel into the extractor, whereas electrons photoexcited from $e_1$ to $e_3$ has a much higher probability to tunnel into the extractor since the $e_3$ envelope wave function is mostly localized in the extractor first QW.

As seen in Fig. 6, the photocurrent spectrum peaked at 23.6 meV (5.7 THz) is well fitted by a Lorentzian lineshape with a full width at half maximum (FWHM) of ΔE/E = 24%. This Lorentzian lineshape of the spectra instead of a Gaussian lineshape attests to the absence of inhomogeneous fluctuations in the grown sample and the reduced linewidth, 5.6 meV, is an indication of its high structural quality. At 20 K, the photocurrent signal persists despite an increase of the noise level. The photocurrent disappears at 30 K. One of the QCD operating temperature limitation is the energy difference between the QCD Fermi level ($E_F + E_1$) and the lowest extractor states energy $E_6$, very small in a THz QCD. From 10 K to 30 K in our devices, this energy is only 3.5 meV due to a high doping level. At 30 K, a parasite channel from $e_1$ to $e_6$ is activated. To confirm this assumption, an analysis of $R_{(V = -0.05\ V)}.A$ in Fig. S3 results in an activation energy $E_a$ of 4 meV at 30 K in good agreement with $E_6 - (E_F + E_1) = 3.5$ meV. Noteworthy, optical characterizations done on a similar GaN THz QCD with a lower doping level ($N \sim 2 \times 10^{11}\ cm^{-2}$) didn't show any photocurrent.

Mesas with 16 μm and 20 μm period gratings showed similar spectra but with a lower peak photoresponse (Cf. Fig. S1). Features at 10 THz probably come from noise as they did not appear in other taken spectra.

We have calibrated the QCD responsivity by measuring the glow-bar source output power with a THz power-meter (TK TeraHertz Absolute Power Meter System). The QCD responsivity at 5.7 THz is estimated to be 0.1 mA/W at 10 K. Responsivity is defined[44] as $R = \frac{q}{E_{31}} \eta \frac{p_e}{N_{QW} p_c}$. $p_e$ is the escape probability, $p_c$ is the capture probability, $E_{31}$ is the operating energy transition, $\eta$ the absorption efficiency is roughly the same for QWIP and QCD from the same spectral range and material system, $N_{QW}$ is the number of periods and $q$ is the elementary charge. The reduced responsivity value (0.1 mA/W), which is significantly lower than the one of previous GaN 10 THz QWIP[27] (10 mA/W), is due to the large capture probability $p_c$ and small escape probability $p_e$ inherent to the QCD nature compared to their QWIP counterpart.

One possibility for increasing the responsivity and maximum operating temperature is to enhance light coupling into a very thin active region with metamaterials. This method achieved an improvement of an arsenide-based THz QWIP performances from a responsivity 0.1 A/W and a maximum operating temperature of 20 K in mesa configuration to 0.25 A/W and a maximum working temperature of 60 K at 5.2 THz when coupling incident light with metamaterial.[45]

In conclusion, we have demonstrated a GaN/AlGaN QCD showing a photoresponse peaked at 5.7 THz (23.6 meV). The photocurrent persists up to 20 K. This GaN/AlGaN quantum cascade device with an active region thickness in the micrometer range has been grown by MOVCD and it achieves efficient vertical transport and operation in the THz range. This thick sample presents a Lorentzian lineshape of the photocurrent with a reduced linewidth of 5.6 meV (corresponding to a record value in of FWHM of $\Delta E/E = 24\%$ for THz range ISB transitions in polar *c*-plane wurtzite III-nitrides nanostructures) and an agreement between grown devices I-V characteristics and theoretical conduction band structure. Those are optical and electrical signatures of a high structural quality of the grown III-nitride material. This result attests to the possibility to synthetize monocrystalline quality III-nitrides QWs potentially satisfying requirement for ISB THz emitters in present-day. This opens prospects for exploring future developments of nitride-based THz QCLs.

See the supplementary material for the photocurrent of GaN/AlGaN QCD at 10 K with 10 μm, 16 μm and 20 μm period grating in Fig. S1, the simulated ISB absorption intensity as a function of periods grating width with 50 % filling factor in Fig. S2 and the Arrhenius plot of R.A versus the inverse temperature in Fig. S3.

**Acknowledgments**

The authors acknowledge technical support of J.-M Manceau, L. Largeau, D. Bouville, F. Maillard and A. Harouri at the Centre de Nanosciences et de Nanotechnologies and this work was supported by French technology facility network RENATECH and the French National Research Agency (ANR) through the projects OptoTeraGaN (ANR-15-CE24-0002) and the "Investissements d'Avenir" program GaNeX (ANR-11-LABX-0014). A.V. Babichev acknowledges the support of the Russian Science Foundation (project no. 20-79-10285) for the part of XRD analysis.

## Figures caption

FIG. 1: Sample description and conduction-band profile and squared moduli of the envelope functions of one period followed by the next period active well of the GaN/AlGaN QCD considered in this study. The layer sequence for one period is **3.9**/ 3.9/ $\overline{7.7}$/ **3.1**/ $\underline{3.1}$/ $\overline{1.2}$/ **3.1**/ $\underline{3.1}$/ $\overline{1.6}$/ **3.1**/ $\underline{3.1}$/ $\overline{2.0}$/ **3.1**/ $\underline{3.1}$/ $\overline{2.7}$ nm. $Al_{0.1}Ga_{0.9}N$ barriers are indicated in bold font. $Al_{0.05}Ga_{0.95}N$ step layers are indicated by the upperline. The doped GaN layers, indicated above by the underline, are n-doped with n = 7 × $10^{17}$ $cm^{-3}$.

FIG. 2: $2\theta - \omega$ scan around the (0002) reflection of the GaN/AlGaN QCD structure. The blue (green) curve correspond to the raw measurement (fitting) Based on the satellite peaks positions, the period thickness was deduced to be 46.94 nm, equal to the nominal value within a 2 % deviation. The best fit to the experimental measurement is achieved for an Al concentration of 4.5 % and 9 % which is close to the nominal values

FIG. 3: AFM image of the as-grown GaN QCD with a rms roughness of ~0.6 nm over an area of 5 × 5 $\mu m^2$.

FIG. 4: Current-voltage characteristics of the 500 × 600 $\mu m^2$ GaN/AlGaN QCD from 10 K to 80 K.

FIG. 5: Current-voltage characteristics of the 500 × 600 µm² GaN/AlGaN QCD at 10 K. The inset shows the conduction band and squared envelope functions of one QCD period with an applied bias of -0.73 V.

FIG. 6: Photocurrent (PC) of a 500 × 600 µm² GaN/AlGaN QCD at 10 K and 20 K with Lorentzian fit from 2 to 15 THz.

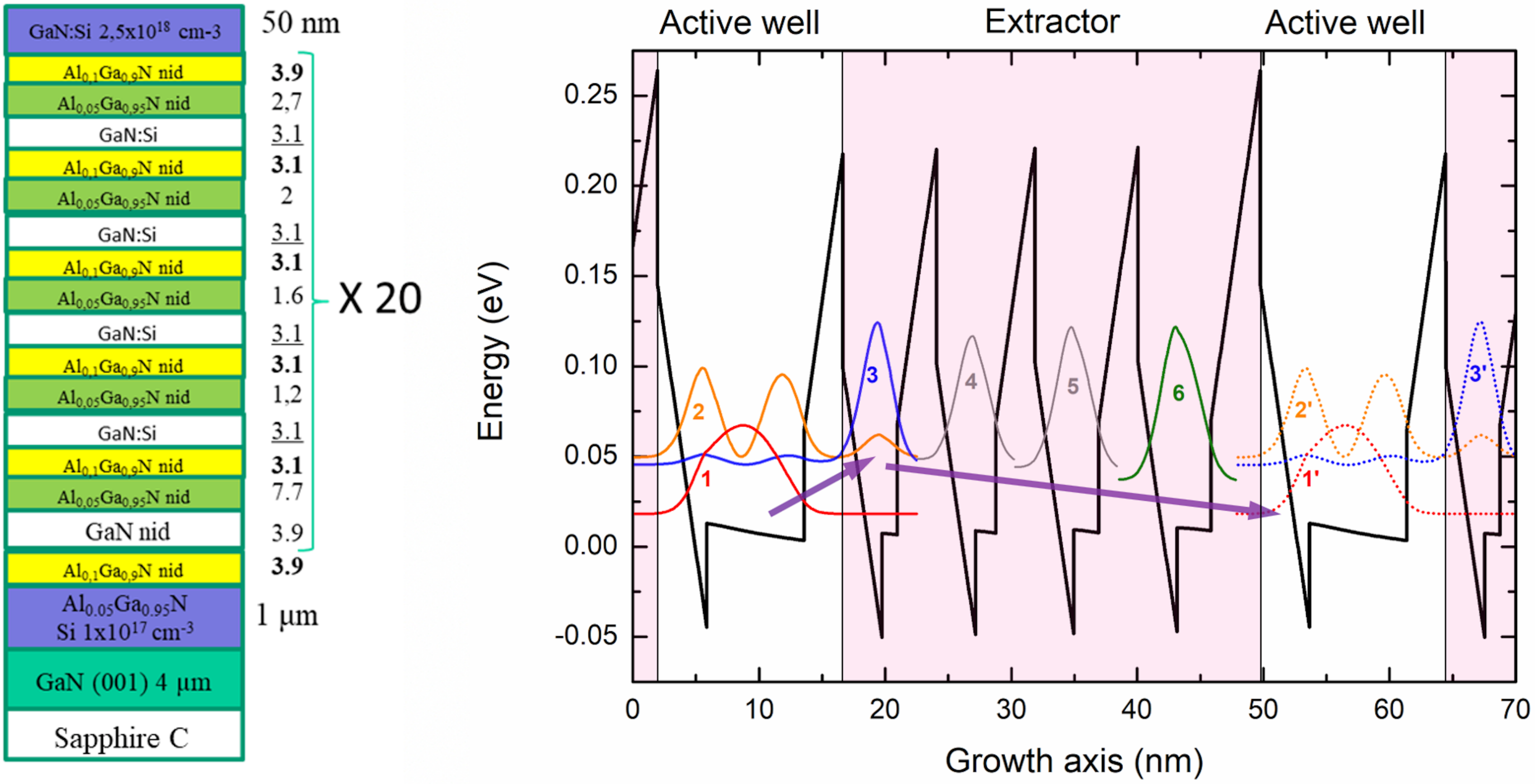

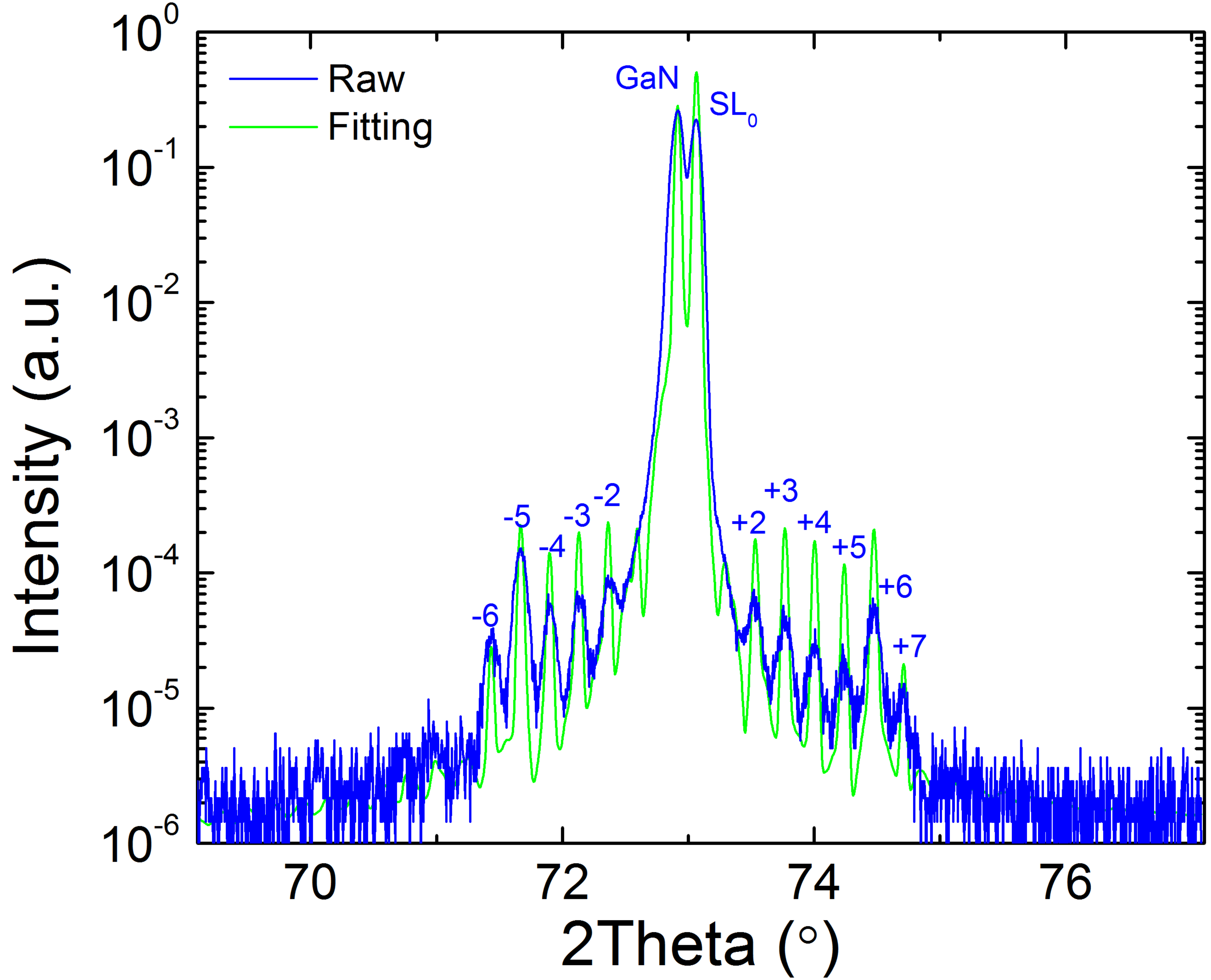

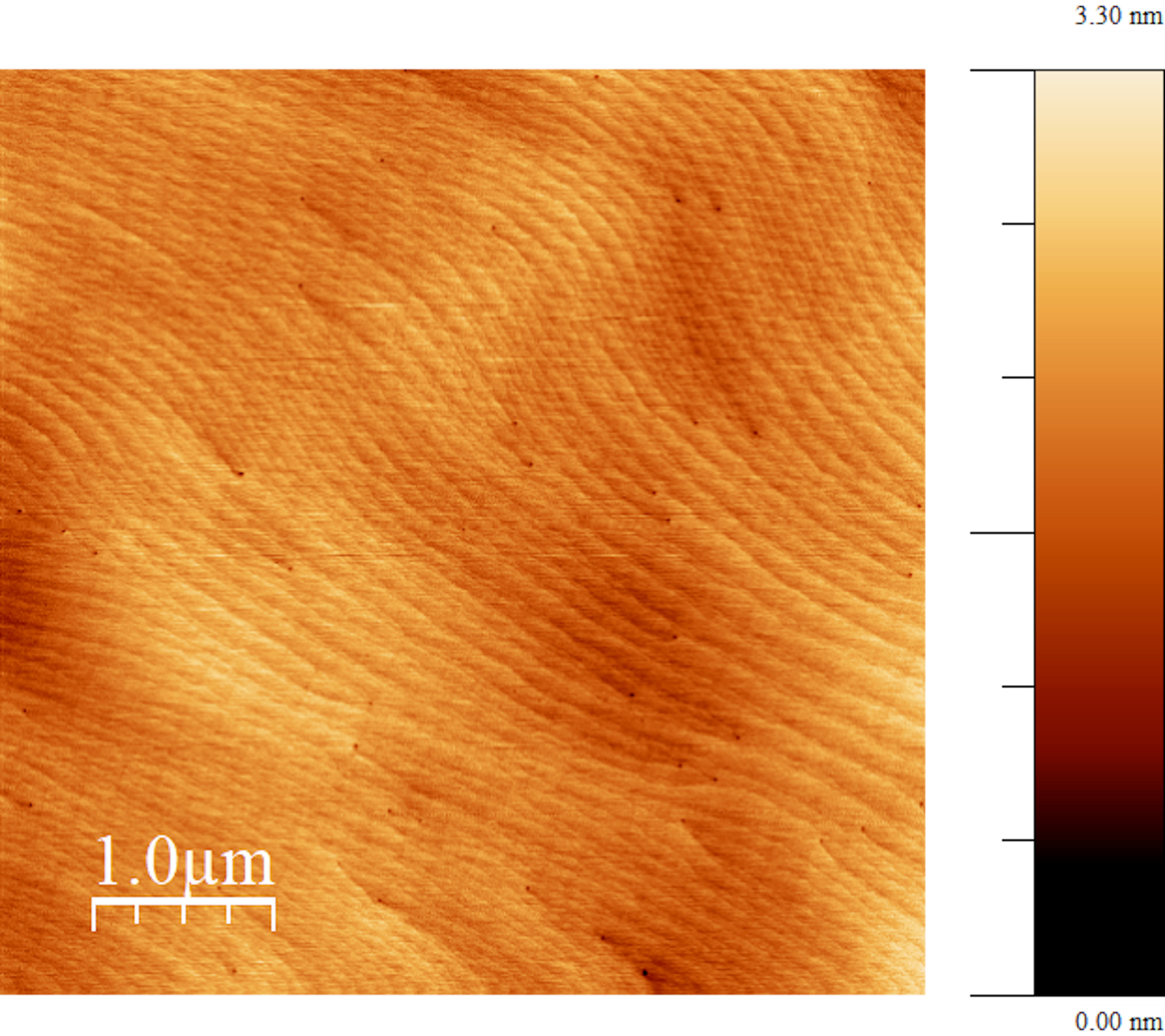

3.30 nm
1.0µm
0.00 nm

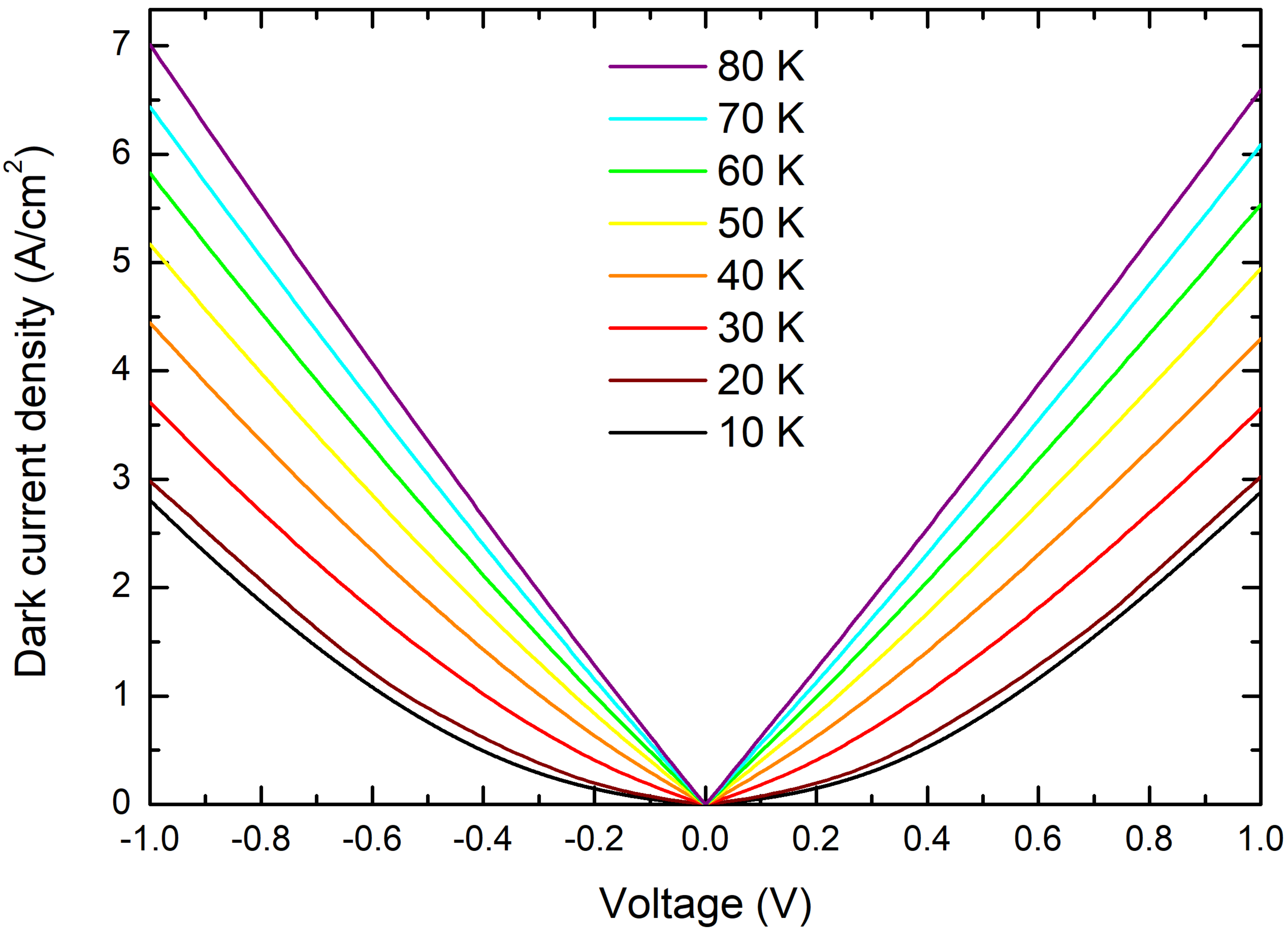

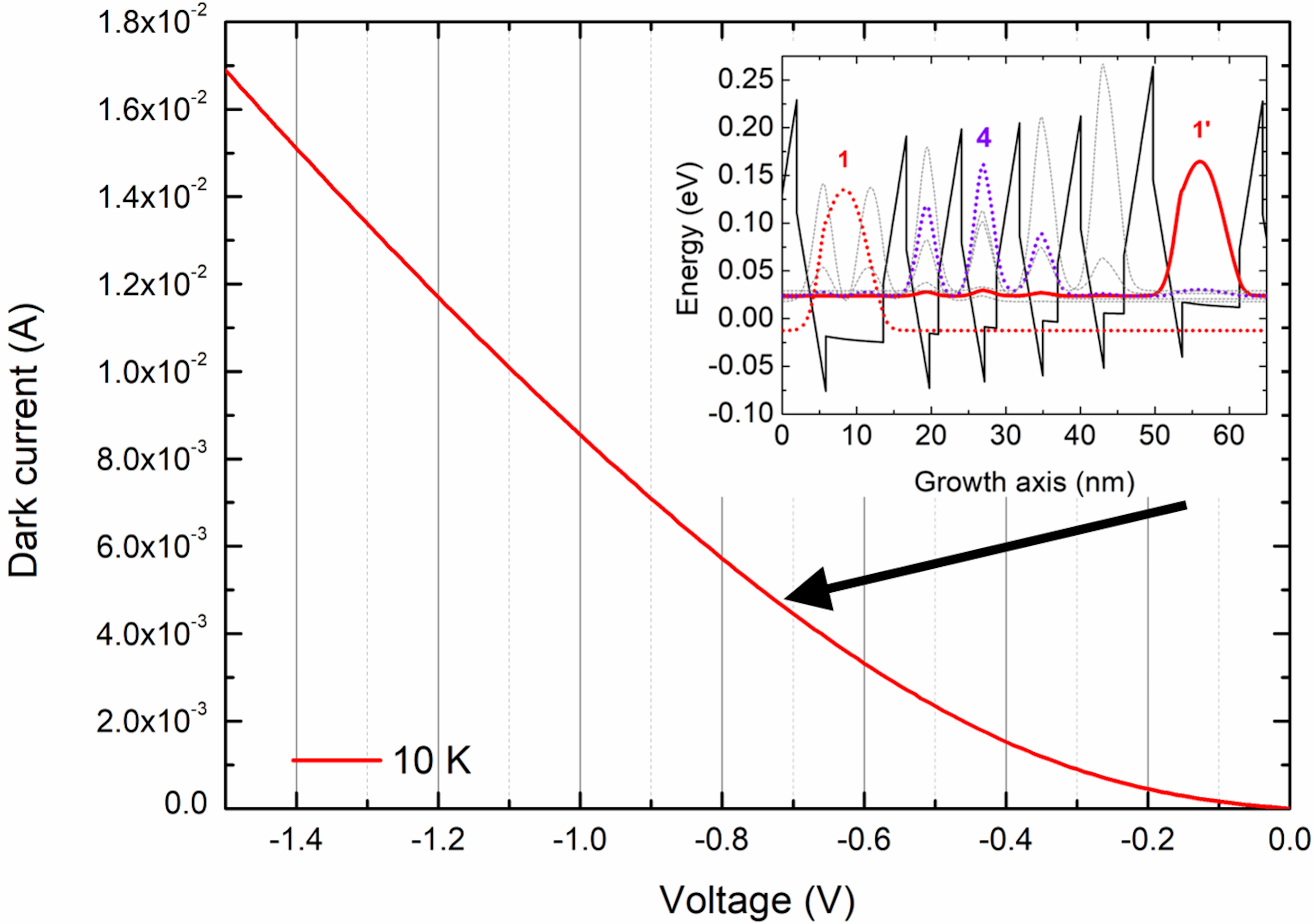

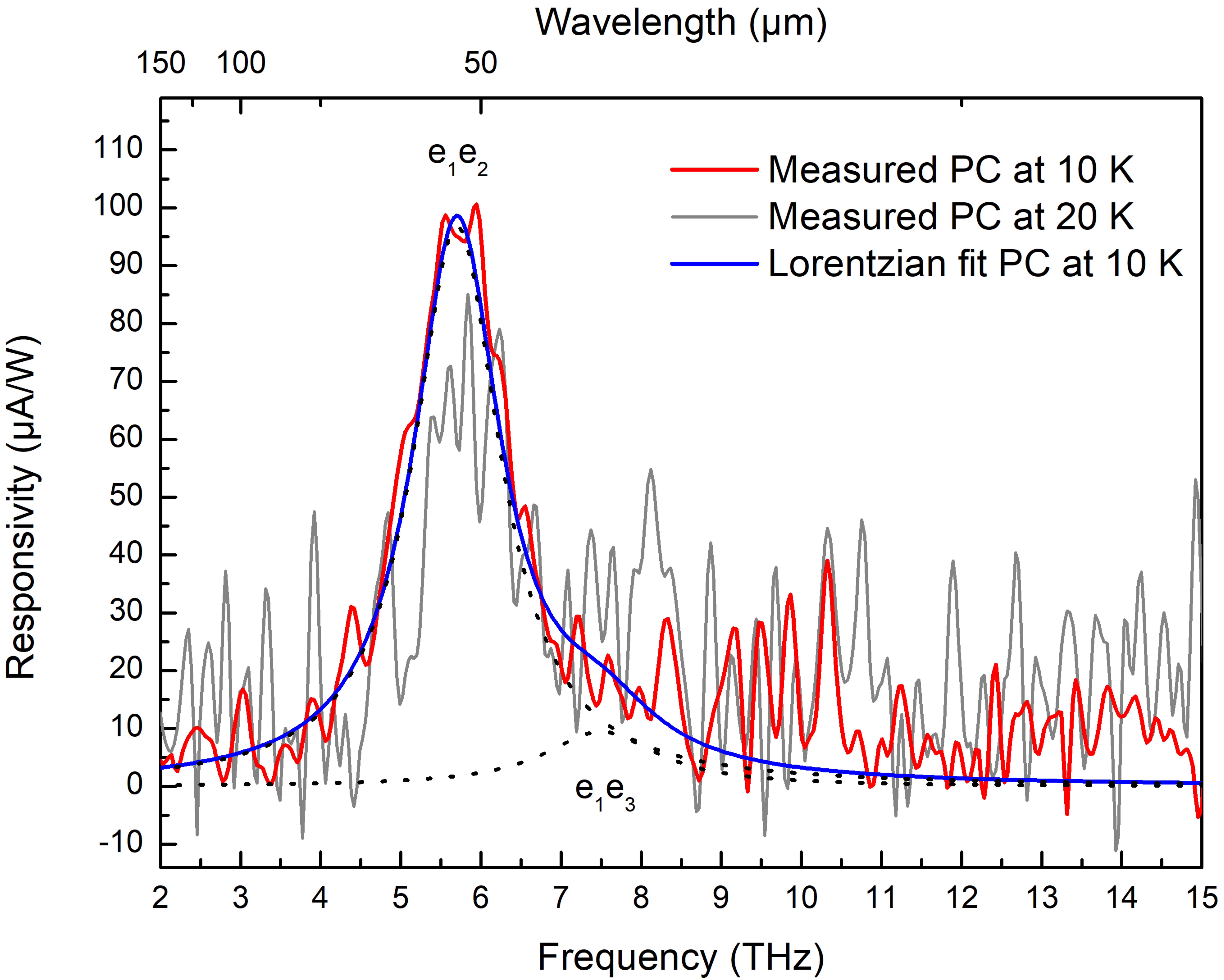